\documentclass[paper,10pt]{article}
\usepackage[dvips]{graphicx}
\begin{document}
\title{Discontinuous Phase Transition in an Exactly Solvable One-Dimensional Creation-Annihilation System}
\author{F H Jafarpour\footnote{Corresponding author's e-mail:farhad@ipm.ir} \, and B Ghavami \\ \\
{\small Bu-Ali Sina University, Physics Department, Hamadan, Iran}}
\maketitle
\begin{abstract}
An exactly solvable reaction-diffusion model consisting of
first-class particles in the presence of a single second-class
particle is introduced on a one-dimensional lattice with periodic
boundary condition. The number of first-class particles can be
changed due to creation and annihilation reactions. It is shown that
the system undergoes a discontinuous phase transition in contrast to
the case where the density of the second-class particles is finite
and the phase transition is continuous.
\end{abstract}
\maketitle
\begin{figure}
  \includegraphics{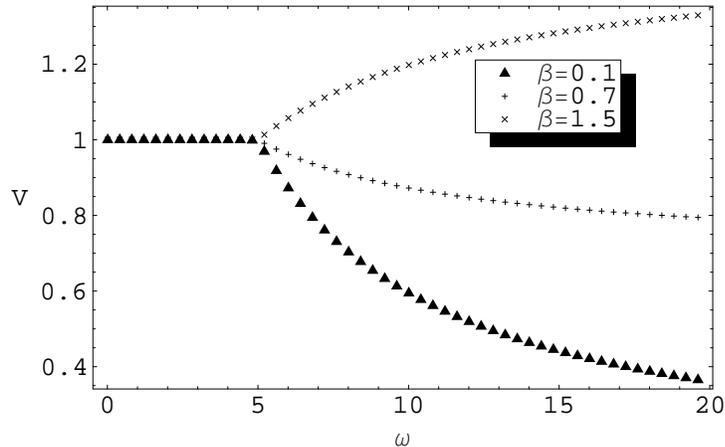}\\
  \caption{The mean speed of the second-class particle $V$ as a function
  of $\omega$ for three values of $\beta$. The length of the system is
  $L=100$ and we have chosen $\alpha=6$.}
  \label{fig}
\end{figure}

 %
One of the most important characteristics of non-equilibrium driven
systems is that their steady state consist of current of particles
or energy. One-dimensional reaction-diffusion models are examples of
such systems which have attracted much attention during last decade
\cite{sz,sch}. Phase transition and shock formation in these systems
are some of their interesting collecting behaviors. These systems
have also many applications in different fields of physics and
biology. During recent years different models of this type have been
studied widely and interesting results have been obtained. The
Asymmetric Simple Exclusion Process (ASEP) is a well known example.
In this exactly solvable model, which is defined on an open discrete
lattice, particle are injected from the left boundary and extracted
from the right boundary while hopping on the lattice to the left and
to the right randomly. This model has been shown to exhibit
nontrivial steady-state phenomena such as phase transitions and
shock formation \cite{dehp}. In order to study the steady state
properties of these shocks different models have been proposed. It
should be noted that the ASEP is not the only one-dimensional
out-of-equilibrium system which exhibits shocks. It has been shown
that there are three families of two-states models in which a
factorized shock measure is invariant under the time evolution if
some constraint on the microscopic reaction rates are fulfilled
\cite{kjs}. In \cite{jm} the authors have shown that the
same phenomenon might also be observed in three-states systems.\\
In this paper we study an exactly solvable three-states model with
non-conserving dynamics on a discrete lattice with a ring geometry.
Our model belongs to the class of non-conserving driven-diffusive
systems where attachment and detachment of particles are allowed .
The study of such models which are variants of the ASEP is motivated
by the biological transport processes in living systems
\cite{pff,lkn} and the denaturation transition in DNA \cite{ps,kmp}.
Since our model is based on a newly introduced model in \cite{jg},
we will first briefly review the main concepts and results of this
model. In \cite{jg} a non-equilibrium three-species system is
introduced on a lattice with periodic boundary condition consisting
of the following reaction processes
\begin{equation}
\label{Process}
\begin{array}{c}
A \emptyset \longrightarrow \emptyset A \;\;\; \mbox{with rate $\alpha$} \\ %%%%%%
\emptyset B \longrightarrow B \emptyset  \;\;\; \mbox{with rate $\beta$} \\ %%%%%%
A B \longrightarrow B A \;\;\; \mbox{with rate 1}  \\ %%%%%%
A \emptyset \longrightarrow \emptyset \emptyset \;\;\; \mbox{with rate $\lambda$} \\ %%%%%%
\emptyset \emptyset \longrightarrow  A \emptyset   \;\;\; \mbox{with rate $\lambda '$}. %%%%%
\end{array}
\end{equation}
As can be seen the number of $A$ particles (first-class particles)
is not conserved. In contrast, the number of $B$ particles
(second-class particles) is conserved since they only diffuse. It is
assumed that in a system with at least one empty site one has finite
number of second-class particles with the density $\rho_{B}$ in the
presence of the first-class particles with fluctuating density. It
has been shown that in this case a continuous phase transition takes
place if the order parameter of the system is taken to be the
density of empty sites in the system $\rho_{E}$. By taking
$\alpha=\beta=1$ and defining $\omega:=\frac{\lambda}{\lambda '}$ it
turns out that $\rho_{E}$ is zero for $\omega < \omega_c$ while it
changes linearly as $\rho_{E}=\frac{\omega}{1+\omega}-\rho_{B}$ for
$\omega > \omega_c$ in which $\omega_c=\frac{\rho_{B}}{1-\rho_{B}}$.
The current of the second-class is always constant while the
particle current of the first-class particles is given by different
expressions in each phase. For the case $\alpha \neq 1$ and
$\beta=1$ the transition point is obtained to be
$\omega_c=\frac{\rho_{B}+\alpha-1}{1-\rho_{B}}$. The density of the
empty sites is zero below the transition point while it is given by
$\rho_{E}=\frac{\omega}{1+\omega}-\frac{\omega\rho_{B}}{1+\omega-\alpha}$
above this point. For $\rho_{B}\neq 0$ the transition is still
continuous.\\
In present paper we assume that there exists only a single
second-class particle in the system which means their density goes
to zero in the thermodynamic limit. Second-class or tagged particles
are usually introduced to study the dynamical properties of the
shocks in one-dimensional driven-diffusive systems; however, one of
our major motivations for studying such limiting case is to
investigate its effects on the critical behavior of the system and
compare it with the previous case in which the density of the
second-class particles is non-zero in the thermodynamic limit. As we
will see considering this limiting case changes the nature of phase
transition from a continuous into a discontinuous one. As far as we
know such observation had not been reported before. Apart from the
vast applicability of such models in different fields of science (as
mentioned above), classification of one-dimensional driven-diffusive
models which are exactly solvable using the Matrix Product Formalism
(MPF) has been of great interests for people in this field (for a
recent review see \cite{be}). As we will see the model is still
exactly solvable using the MPF even in the limiting case
$\rho_B \rightarrow 0$.\\
In the following we define $\omega:=\frac{\lambda}{\lambda '}$ and
apply the MPF \cite{dehp} to find the partition function of the
system. According to the MPF the stationary probability distribution
function of any configuration $\mathcal{C}$ of the system of length
$L+1$ with a single second-class particle at the site $L+1$ is given
by
\begin{equation}
\label{SPDF} P(\mathcal{C})=\frac{1}{\mathcal{Z}}
Tr[(\prod_{i=1}^{L}{\bf X}_i){\bf B}]
\end{equation}
in which ${\bf X}_i={\bf E}$ if the site $i$ is empty otherwise
${\bf X}_i={\bf A}$. The normalization factor $\mathcal{Z}$ in the
denominator of (\ref{SPDF}) will be called the partition function of
the system. By applying the standard MPF the quadratic algebra of
the model is obtained to be \cite{jg}
\begin{equation}
\label{ALG1}
\begin{array}{l}
{\bf A} {\bf B}={\bf A}+{\bf B} \\
{\bf A} {\bf E}=\frac{1}{\alpha}{\bf E}\\
{\bf E} {\bf B}=\frac{1}{\beta}{\bf E}\\
{\bf E}^2=\frac{\omega}{\alpha}{\bf E}.
\end{array}
\end{equation}
By defining ${\bf E}=\frac{\omega}{\alpha}\vert V \rangle \langle W
\vert$ in which $\langle W \vert V \rangle=1$ one finds from
(\ref{ALG1})
\begin{equation}
\label{ALG2}
\begin{array}{l}
{\bf A} {\bf B}={\bf A}+{\bf B} \\
{\bf A}\vert V \rangle= \frac{1}{\alpha}\vert V \rangle\\
\langle W \vert {\bf B}=\frac{1}{\beta}\langle W \vert.
\end{array}
\end{equation}
This quadratic algebra has an infinite-dimensional representation
given by the following matrices and vectors
\begin{equation}
\label{REP}
\begin{array}{c}
{\bf A}=\left(
    \begin{array}{ccccc}
      \frac{1}{\alpha} & a & 0 & 0 & \cdots \\
      0 & 1 & 1 & 0 &  \\
      0 & 0 & 1 & 1 &  \\
      0 & 0 & 0 & 1 &  \\
      \vdots &   &   &   & \ddots \\
    \end{array}
  \right), {\bf B}=\left(
    \begin{array}{ccccc}
      \frac{1}{\beta} & 0 & 0 & 0 & \cdots \\
      a & 1 & 0 & 0 &  \\
      0 & 1 & 1 & 0 &  \\
      0 & 0 & 1 & 1 &  \\
      \vdots &   &   &   & \ddots \\
    \end{array}
  \right),\\
\vert V \rangle=\left(
                                    \begin{array}{c}
                                      1 \\
                                      0 \\
                                      0 \\
                                      0 \\
                                      \vdots \\
                                    \end{array}
                                  \right),
\langle W \vert =\left(
                                    \begin{array}{ccccc}
                                      1&0&0&0&\cdots \\
                                    \end{array}
                                  \right)
   \end{array}
\end{equation}
in which $a^2=\frac{\alpha+\beta-1}{\alpha\beta}$. Since the
stationary state of the system without vacancies is trivial, we
consider the partition function of the system with at least one
empty site which is defined by
\begin{equation}
\label{PF1} \mathcal{Z}=Tr[({\bf A}+{\bf E})^L{\bf B}]-Tr[{\bf
A}^L{\bf B}].
\end{equation}
Using (\ref{ALG2}-\ref{PF1}) and after some straightforward
calculations we find the following exact expression for the
partition function of the system
\begin{equation}
\label{PF2}
\mathcal{Z}=\frac{\beta+\omega}{\beta(1+\omega-\alpha)}(\frac{1+\omega}{\alpha})^L-
\frac{1}{1-\alpha}(\frac{1}{\alpha})^L+\frac{\omega(\alpha+\beta-1)}{\beta(1+\omega-\alpha)(1-\alpha)}.
\end{equation}
For $\alpha < 1$ there is no phase transition. Assuming $\alpha > 1$
one simply finds the following expressions for the partition
function of the system in the large $L$ limit
\begin{equation}
\label{PF3}
\mathcal{Z}\cong\left\{
\begin{array}{ll}
\frac{\beta+\omega}{\beta(\beta+\omega-\alpha)}(\frac{1+\omega}{\alpha})^L&
\mbox{for $\omega>\alpha-1$}\\ \\
\frac{\omega(\alpha+\beta-1)}{\beta(1+\omega-\alpha)(1-\alpha)}&
\mbox{for $\omega<\alpha-1$}.
\end{array}
\right.
\end{equation}
At the transition point the partition function of the system grows
like $\mathcal{O}(L)$. Taking the density of the empty sites on the
lattice given by
\begin{equation}
\label{DofE1} \rho_{E}=\lim_{L\rightarrow
\infty}\frac{\omega}{L}\frac{\partial}{\partial \omega}\ln {\cal Z}
\end{equation}
as the order parameter of the system, we find using (\ref{PF3}) that
\begin{equation}
\label{DofE2} \rho_{E}=\left\{
\begin{array}{ll}
\frac{\omega}{1+\omega}& \mbox{for $\omega>\alpha-1$}\\ \\
0& \mbox{for $\omega<\alpha-1$}.
\end{array}
\right.
\end{equation}
At the transition point the density of the empty sites is obtained
to be $\rho_{E}=\frac{\omega}{2(1+\omega)}$. We should note that
density of the empty sites for $\omega<\alpha-1$ drops to zero as
$\rho_{E}\propto \mathcal{O}(\frac{1}{L})$. Discontinuous changes of
the density of the empty sites $\rho_{E}$ in the thermodynamic limit
indicates a first-order phase transition in the system. As we
mentioned earlier, in the case where the number of the second-class
particles on the lattice is finite the density of the empty sites
$\rho_{E}$ changed continuously over the transition point
\cite{jg}.\\
In order to study the nature of the first-order phase transition one
can apply the Yang-Lee theory. Recently it has been shown that the
classical Yang-Lee theory can be applied to the out-of-equilibrium
systems to study their phase transitions (for a review see
\cite{bdl}). We have calculated the line of the Yang-Lee zeros for
our model in the complex-$\omega$ plan and found that they lie on a
circle of radius $\alpha$. The center of this circle is at $(-1,0)$
and intersects the real-$\omega$ axis at $Re(\omega)=\alpha-1$ at an
angle $\frac{\pi}{2}$ which again implies a first-order phase
transition at the transition point. The density of the zeros has also
been found to be a constant all over the circle.\\
It is also interesting to calculate the density profile of the first
class particles on the ring, as seen by the second-class particle,
using the MPF. For a system with at least one empty site it is given
by
\begin{equation}
\label{DPA1} \rho_{A}(i)=\frac{1}{\cal
Z}(Tr[(A+E)^{i}A(A+E)^{L-i-1}B]-Tr[A^{L}B]) \;\;\;\; 0 \le i \le
L-1.
\end{equation}
It turns out that (\ref{DPA1}) can be calculated exactly using
(\ref{REP}) and here are the results in the large $L$ limit
\begin{equation}
\label{DPA2} \rho_{A}(i)\cong\left\{
\begin{array}{ll}
\frac{1}{1+\omega}+\frac{\omega(\alpha+\beta-1)}{\alpha(\beta+\omega)}e^{\frac{i-L}{\xi}}&
\mbox{for $\omega>\alpha-1$}\\ \\
1-\frac{\alpha-1}{\alpha}e^{-\frac{i}{\xi}}& \mbox{for
$\omega<\alpha-1$}\\
\\ \frac{1}{1+\omega}+\frac{\omega}{1+\omega}(\frac{i}{L})&\mbox{for
$\omega=\alpha-1$}
\end{array}
\right.
\end{equation}
in which the correlation length is given by $\xi=\vert
\ln(\frac{1+\omega}{\alpha})\vert^{-1}$. For $\omega>\alpha-1$ the
lattice is filled by first-class particles of density
$\frac{1}{1+\omega}$ except just in front of the second-class
particle where it increases exponentially to $1$. In this phase the
density of empty sites is $\frac{\omega}{1+\omega}$. For
$\omega<\alpha-1$ the density of first-class particles increases
exponentially from $\frac{1}{\alpha}$ to $1$ in the bulk of the
lattice. The density of empty sites in this phase is nearly zero in
the thermodynamic limit. As can be seen at the transition point the
density profile of the particles is linear. This is a sign for a
shock however since the number of first-class particles is not a
conserved quantity the shock position fluctuates and therefore the
resulting profile is linear. This phenomenon has also been observed
in the ASEP with open boundaries on the first-order phase transition
line where the injection and extraction rates become equal and
smaller than one-half. The sock picture will be more clear by
calculating the connected two-point function of first-class
particles. Straightforward calculations result in the following
exact expression which is valid for $i \le j$ and large system
length
\begin{equation}
\label{CTPF} \begin{array}{ccc}\langle
\rho_{A}(i)\rho_{A}(j)\rangle_{c} & := & \langle
\rho_{A}(i)\rho_{A}(j)\rangle-\langle \rho_{A}(i)\rangle \langle
\rho_{A}(j)\rangle \\ \\ & \cong & -(\frac{1}{1+\omega}-\langle
\rho_{A}(i)\rangle)(1-\langle \rho_{A}(j)\rangle).
\end{array}
\end{equation}
We have also calculated the current of the first-class particles
$J_A$ in the steady state. In the large $L$ limit we have found that
the current of the first-class particles does not depend on $\beta$
and is given by
\begin{equation}
J_A=\left\{
\begin{array}{ll}
\frac{\alpha\omega}{(1+\omega)^2}& \mbox{for $\omega>\alpha-1$}\\ \\
0& \mbox{for $\omega<\alpha-1$}\\ \\
\frac{\omega}{2(1+\omega)}& \mbox{for $\omega=\alpha-1$}.
\end{array}
\right.
\end{equation}
On the other hand, the mean speed of the second-class particle
defined as
\begin{equation}
V=\frac{1}{\mathcal{Z}}(\beta
Tr[(A+E)^{L-1}EB]+Tr[(A+E)^{L-1}AB]-Tr[A^LB])
\end{equation}
can also be calculated exactly. It turns out that $V$ is given by
the following exact expression in the thermodynamic limit
\begin{equation}
V=\left\{
\begin{array}{ll}
\frac{\alpha\omega+\beta((1+\omega)^2-\alpha\omega)}{(1+\omega)(\beta+\omega)}& \mbox{for $\omega > \alpha-1$}\\ \\
1& \mbox{for $\omega \leq \alpha-1$}.
\end{array}
\right.
\end{equation}
In Figure \ref{fig} we have plotted $V$ as a function of $\omega$
for three different values of $\beta$ on a lattice of length
$L=100$.
For $\omega < \alpha-1$ the speed of the second-class particle is
equal to one and does not depend on $\omega$ while for $\omega <
\alpha-1$, above the transition point, for $\beta < 1$ ($\beta > 1$)
the speed of the second-class particle is a decreasing (increasing)
function of $\omega$. For $\beta=1$ the speed of the second-class
particle is always equal to unity.\\
The model studied in this paper consists of a single second-class
particle in the presence of first-class particles with fluctuating
density because of creation and annihilation of them. Comparing out
results with those obtained in \cite{jg} one should note that the
order of the phase transition is changed from two to one when the
density of the second-class particles goes to zero. The physical
explanation for such a macroscopic change can be as follows: for the
two cases $\rho_B = 0$ and $\rho_B \neq 0$ and below the critical
point there are empty sites on the ring; however, their density goes
to zero in the thermodynamic limit. Nevertheless, it is less
probable to find configurations of type $A \emptyset$ in the case
$\rho_B\neq 0$ (in comparison to the case $\rho_B=0$) because in
this case part of the system is occupied by the second-class
particles. Therefore as we increase $\omega$ above the critical
point it is more probable to create many empty sites in the case
$\rho_B=0$ than the case $\rho_B \neq 0$ because as explained above
we have more configurations of type $A\emptyset$ in this case which
result in the configuration $\emptyset \emptyset$. This means that
as we increase $\omega$ above the critical point we expect many
empty sites to be created at once in the case $\rho_B=0$ in contrast
to the case $\rho_B \neq 0$ where they are being created slowly.\\
It is also interesting to compare our results with those obtained in
\cite{lkp} where the same model as (\ref{Process}) has been
considered except it does not contain the creation and annihilation
of the first-class particles. It has been shown that the model in
this case has two phases: a condensate phase and a fluid phase.
Obviously in our model by fixing the number of the first-class
particles the phase in which $\rho_{E}=0$ will be destroyed;
however, a condensate phase emerges in which the density profile of
the first-class particles is no longer linear but an step-function
and this is in quite agreement with the results in \cite{lkp}.\\
The MPF enables us to solve some of the one-dimensional
driven-diffusive systems exactly; however, by looking at these
models we realize that they might have similar quadratic algebras.
This means that a quadratic algebra can describe different models
with different physical properties and critical behaviors. Now the
question is that whether or not other three-states models defined on
a ring geometry can be described by (\ref{ALG1}). In fact we have
found that there is a family of such models which are exactly
solvable and share the same quadratic algebra \cite{jk}. So far two
members of this family are introduced and studied in \cite{jg} and
\cite{eklm}.

\end{document}